\documentclass{ws-procs975x65}            

\def\be{\begin{eqnarray}}\def\ee{\end{eqnarray}}
\def\lsim{\mathrel{\rlap{\lower3pt\hbox{\hskip1pt$\sim$}}
     \raise1pt\hbox{$<$}}} 
\def\gsim{\mathrel{\rlap{\lower3pt\hbox{\hskip1pt$\sim$}}
     \raise1pt\hbox{$>$}}} 
\def\le{ \begin{array}{ll}}\def\re{\end{array}}

\def\lear{ \left( \begin{array}{cc}}\def\rear{\end{array} \right)}

\def\le{ \left( \begin{array}{cc}}\def\re{\end{array} \right)}

\def\bi{\bibitem}

\def\la{\langle}\def\ra{\rangle}
\def\del{\partial}
\begin{document}
\title{The Proton Mass and Scale-Invariant Hidden Local Symmetry
\\ for Compressed Baryonic Matter}

\author{Mannque Rho $^*$}

\address{Institut de Physique Th\'eorique, CEA Saclay, 91191 Gif-sur-Yvette c\'edex, France\\
$^*$E-mail: mannque.rho@cea.fr}

\begin{abstract}
I discuss how to access dense baryonic matter of compact stars by combining hidden local symmetry (HLS) of light-quark vector mesons with spontaneously broken  scale invariance of a (pseudo) Nambu-Goldstone boson, dilaton,  in a description that parallels the approach to dilatonic Higgs. Some of the surprising observations are that the bulk of proton mass is not Nambu-Goldstonian, parity doubling emerges at high density and the EoS of baryonic matter can be soft enough for heavy-ion processes at low density and stiff enough at high density for $\sim 2$ solar mass neutron stars.
\end{abstract}


\bodymatter
%

\section{Introduction}
Every speaker in this meeting, perhaps with one glaring exception, that is, myself,  has so far been, and will be, talking about the Higgs boson or Higgs-like scalars and matters related to mass generation and above all what's beyond the Standard Model. But I am not in the field and that's not what I will talk about. So why am I invited by Koichi to give this talk and what is it all about ?

Here is why and what.

Whatever the true mechanism for the origin of the mass of the visible Universe, hotly debated in this meeting, may be, the Higgs boson may very well account for the mass of the constituents of the proton, say, quarks, but I will suggest that its presence {\it by itself} cannot explain the bulk of the proton mass. I will present this issue using what's being observed of massive compact stars.

Although the physics involved may look quite different, it seems that there are quite a few things in common between what's being discussed in this meeting and what I will talk about, i.e., dense  baryonic matter that is on the verge of collapsing into a black hole. In my view it has a lot to do with light-quark vector bosons endowed with hidden local symmetry, the idea that initially germinated here in Nagoya and a scalar dilaton associated with spontaneously broken scale symmetry, which is one of the main goings-on in Nagoya, the techni-dilaton~\cite{koichi}.

In describing this matter in my talk I will be a bit daring. I will do this without dragging in my close collaborators who have been working with me on the matter since some time because I don't want to  involve them in my discussions, some of which could be too wild and above all I don't want them to be held responsible for some far-out ideas I will launch onto in this talk.

My talk is about the origin of the mass of the ``ordinary matter" we see all around us. Now that the fundamental boson considered to be responsible for the origin of the mass of the Universe, the Higgs boson, is discovered, one would think that we now know where the mass of the proton, say,  comes from. Given the precisely measured nucleon (say, proton) mass, the mass of all other visible  matter around us -- this building, molecules, atoms and nuclei -- is almost completely accounted for: More than 99\% of the mass is given by the simple sum of the masses of the nucleons involved. But this simple picture ends there. The mass of the nucleon is not just sum of ``things."  One misses more than 95\% of the proton mass if one simply adds the masses of the QCD ``things" of the nucleon, i.e., quarks, the mass of which is presumably given by the Higgs mechanism, and massless gluons. QCD on lattice, however, does give the proton mass correctly, hence QCD does have in it but it pops out of computers after highly complex and intricate manoeuvering~\cite{wilczek}. But one cannot ``see" it. The first question therefore is where does the proton mass come from? The Higgs mechanism does not provide an obvious answer to this question, so the statement that the origin of the SM mass is the Higgs mechanism cannot be taken to be the whole story. Now simple models like linear $\sigma$ model and equivalently NJL model say that the mass can arise from spontaneous breaking of chiral symmetry -- the Nambu-Goldstone (NG) mechanism -- signalled by the presence of NG bosons, i.e., pions, with the nonzero order parameter, the quark condensate, $\la\bar{q}q\ra\neq 0$. Then the next question is, {\it if the mass is generated dynamically with a nonzero $\la\bar{q}q\ra$ from chiral symmetry, can one not unbreak the symmetry by tweaking the condensate to go to zero by, say, temperature and/or density?}

These are the questions that Gerry Brown and I raised in early 1980's. In the attempt to answer these questions -- which do not require involving directly the microscopic (QCD) degrees of freedom, i.e., quarks and gluons, we came across the hidden local symmetry associated with the light-quark vector mesons $V\equiv (\rho, \omega)$ -- the only effective field theory that we knew had the possibility of having a non-NG hadron with vanishing mass and the spontaneously broken scale symmetry associated with a dilaton,  denoted $\sigma$, to be identified with $f_0(500)$ listed in the particle data booklet. The vectors $V$ and the scalar $\sigma$ have been around in the nuclear physics community since a very long time but what I will present here takes a totally different aspect that though quite different physics-wise, is closely along the line of the thinking currently prevalent in the particle physics community like in this meeting.

In this talk I will focus on density driving the quark condensate to vanish\footnote{For simplicity I will be thinking of the chiral limit in which the pion is massless although confrontation with Nature requires, sometimes quite crucially, the small pion mass to be taken into account.}. This is relevant to the interior of compact stars, where the density is expected to be high enough to be near  chiral restoration. This issue became quite topical and particularly acute with the recent precise measurements of $\sim 2$-solar mass neutron stars.
\section{``Scalar Meson Conundrum" and Vector Mesons in Nuclear Physics}
 In the spirit of effective field theory -- that I will adopt -- a low-mass scalar meson and comparable-mass vector mesons turn out to be absolutely indispensable to understand what's going on both in nuclei and in dense nuclear matter. A scalar meson with a mass around 600 MeV has figured importantly in nuclear physics since a long time. The only scalar mesonic excitation with such a low mass currently listed in the particle data booklet is $f_0(500)$ with a broad width. Despite the observed large width,  when taken, for no good reason other than convenience, as a local bosonic field, it has fairly successfully accounted for the attractive scalar channel in nucleon-nucleon potentials such as the well-known Bonn boson-exchange potential and, more significantly, in the highly popular relativistic mean field (RMF) theories for nuclei, nuclear matter and compact-star matter\footnote{This model consists of the vector mesons $\rho$, $\omega$ and a scalar of mass $\sim 600$ MeV coupled to baryons treated in the mean-field. This model is presumably related to the Landau Fermi-liquid approach to nuclear matter, the reason why the model seems to work well in nuclear physics.}. There has been a long-standing controversy as to whether the scalar excitation with such a large width can be depicted in terms of a local field in phenomenological or effective Lagrangian approaches. If it is a hadronic particle, then what is its QCD structure? Can it be described in quark models, as, say, quarkonium of $(q\bar{q})^m$ complex with $m\geq 1$, a gluonium (glueball) or a mixture thereof? In effective field theories of strong interactions, there is the fourth component of the chiral four-vector in linear sigma model. However it must have a mass $\gsim 1$ GeV to be compatible with the current algebras, so cannot be relevant far below the chiral scale $4\pi f_\pi\sim 1$ GeV. Furthermore what figures in  relativistic mean field (RMF) theories -- which are highly (perhaps too) successful in heavy nuclei and nuclear matter -- must be a chiral-singlet scalar, not the scalar of linear sigma model. If it were the sigma-model scalar with a mass $< 1$ GeV as needed in nuclear phenomenology, it would destabilize nuclear matter due to strong attractive many-body forces. A possible way-out of this difficulty is that the scalar is a low-lying chiral singlet scalar at least in the vicinity of nuclear matter density. However at high density as the system approaches chiral restoration, it should transmute to a $\bar{q}q$ configuration that is expected to dominate at the large $N_c$ limit~\cite{DLFP}.

I suggest that the idea recently put forward by Crewther and Tunstall~\cite{CT} that the scalar is a NG boson or more precisely pseudo-NG (pNG) boson of spontaneously broken scale symmetry can provide a simple and appealing way to resolve the above ``scalar conundrum" and offers a new perspective on various aspects of nuclear physics, including hadronic matter under extreme conditions that have remained poorly understood. The  power of the scheme is that the low-lying scalar, a chiral singlet, can be treated on the same footing as the pseudoscalar pseudo-NG bosons -- the pions and kaons -- in low-energy effective field theory. This is called  ``scale chiral perturbation theory ($s\chi$PT)"~\cite{CT}. I don't think this idea has been confirmed rigorously, but I will take this as an {\it assumption that may be valid in medium} and see how it fares with Nature.

No less crucial for nuclear stability with the low-mass scalar providing the attractive interactions are vector-meson degrees of freedom. They provide the short-range repulsion that counters the scalar attraction at short distances to prevent nuclear matter from collapsing and in addition, control the nuclear tensor forces crucial in nuclear dynamics and, as it turns out, in compact stars.  What figures in RMF theories for repulsive interaction is the isoscalar vector meson, $\omega$, and for the tensor forces important for compact stars, it  is the $\rho$ meson. While both $\rho$ and $\omega$ are essential for both nuclear matter and  dense matter,  their roles are basically different. Endowed with hidden local symmetry (HLS), the $\rho$ meson has a fixed point called ``vector manifestation (VM)" at which the HLS coupling $g$ vanishes as the quark condensate -- order parameter of chiral symmetry --  is tuned to zero, as a consequence of which the $\rho$ mass vainshes~\cite{HY:PR}. In this framework, it is the VM fixed point that baryonic matter will be driven to when the condensate $\la\bar{q}q\ra$ is dialled to zero. I will exploit this property in describing the EoS of star matter, even though the system may not reach the chiral restoration point.

As for the $\omega$, it seems that the $U(2)$ symmetry that puts the $\omega$ in the same multiplet as the $\rho$ works very well in matter-free space. Apart from near degeneracy in mass between them, the vector dominance picture holds very well for both the pion and nucleon EM form factors if one takes into account the infinite towers of the isovector $\rho$'s and isoscalar $\omega$'s -- embedded in 5-D local $U(2)$ symmetry -- as in holographic QCD models~\cite{SS,HRYY}. For this, the infinite tower is essential. However it has been known since Sakurai proposed the idea in 1970's that while the vector dominance with the lowest $\rho$ meson alone worked well for the pion EM form factor,  it failed famously for the nucleon form factors. For the latter, other degrees of freedom such as quark bag~\cite{BRW} were required. This implied among others that the flavor (and local) $U(2)$ symmetry could breakdown, quite drastically, in a vacuum modified by dense matter as I will discuss below.
\section{Implementing Scale Invariance to Hidden Local Symmetry}
An early idea favored by many people in particle and nuclear physics, including Nambu himself (see \cite{schumacher} for a current review), for the origin of $\sim 99\%$ of the proton mass was the spontaneous breaking of chiral symmetry associated with the nearly zero-mass quarks. This idea provided a natural order parameter, the quark condensate $\la\bar{q}q\ra$, for the phase transition that could be probed by ``unbreaking" chiral symmetry  by external disturbances. ``Seeing" dropping mass of the proton by experiments, say, by increased density, would provide a ``smoking gun" signal for the origin of the proton mass.
Unfortunately lattice gauge calculation cannot yet access high density due to the the famous sign problem. Furthermore, there are presently no reliable model-independent theoretical tools available. Experimental efforts are being,  and will be, made in heavy-ion collisions to explore this uncharted area. Up to date  there is no clear indication either for or against the notion that the proton mass can be unsheathed by density or temperature. The only laboratory available at present is the space laboratory and the object is the dense compact star matter.

As I will argue,  the proton mass is not, it turned out, the right object to look at for chiral symmetry. In fact some $\gsim 70$\% of the proton mass seems to remain unaffected by the quark condensate going to zero, thus, contrary to what was generally accepted, unconnected to spontaneously broken chiral symmetry (SBCS). The first suggestion on what to look for in Nature for a signature for chiral symmetry in hadronic matter was made by Gerry Brown with his collaborators~\cite{adami-brown}. His idea was to observe instead dropping $\rho$ mass at high temperature (or density), which ultimately goes to zero at the chiral restoration point (that is, in the chiral limit). He was initially interested in high-temperature cases but I believe the suggestion is more appropriate for high density matter. Heavy-ion collisions producing high temperature matter are too swamped with mundane backgrounds to directly and cleanly probe the vacuum-change caused by temperature.

As mentioned, the only known way in which a zero-mass $\rho$ meson figures {\it naturally} in nuclear dynamics under extreme conditions is in a hidden gauge symmetric setting~\cite{HY:PR}.  It is also in this way that chiral symmetry and scale symmetry could be intricately locked to each other.  This accounts for why Gerry and I advocated -- since many years and more or less isolated from the rest of nuclear physics community -- the approach based on HLS~\cite{HLS,HY:PR} -- and dilaton. We will see that this feature plays an extremely important role in dense medium, particularly in what is called the symmetry energy $E_{sym}$ that is crucial in the EoS of compact-star matter.

Together with the scalar degree of freedom treated below, the $\omega$ meson plays an equally indispensable role in nuclear physics as a whole and is crucial at high density. The $U(2)$ symmetry, while holding well in the matter-free space for the $\omega$  and $\rho$ mesons, is most likely broken badly at high density with the $\omega$ insensitive to the VM fixed point of the $\rho$ meson. I will therefore consider independently $SU(2)$ HLS for the $\rho$ and $U(1)$ HLS for the $\omega$ in treating dense medium. I write the HLS Lagrangian to ${\cal O}(p^4)$ in the expansion in covariant derivatives~\footnote{For convenience, I follow the notation of Ref. \cite{maetal}.} $D_\mu \xi_{R,L}^{} = (\partial_\mu - i V_\mu) \, \xi_{R,L}^{}$, where $V_\mu$ represents the gauge bosons of the HLS as $
V_\mu = \frac{1}{2} \left( g_\omega^{} \omega_\mu^{} + g_\rho^{} \rho_\mu^{} \right)$ with $\rho_\mu=\vec{\tau}\cdot \vec{\rho}_\mu$. The Lagrangian is
\begin{eqnarray}
\mathcal{L}_{\rm HLS} & = &
\mathcal{L}_{(2)}^{\rm HLS} + \mathcal{L}_{(4)}^{\rm HLS} +
\mathcal{L}_{\rm (4) anom}^{\rm HLS} +\mathcal{L}_{\rm mass}+ \cdots \, .
\label{HLS}
\end{eqnarray}
The numerical subscript stands for the power of covariant derivatives. Here $\mathcal{L}_{\rm mass}$ is the quark mass term and the anomalous Wess-Zumino term contains four independent terms. The $V_\mu$ kinetic terms are in the leading order term $\mathcal{L}_{(2)}^{\rm HLS}$.

As stressed in Introduction, the scalar degree of freedom in nuclear interactions has been elusive since a long time. The recent idea, contested by some workers in the field, of Crewther and Tunstall~\cite{CT} that the low mass scalar $f_0(500)$ is a pseudo-Nambu-Goldstone boson -- with the scale symmetry explicitly broken by the trace anomaly -- {\it and} the quark mass -- arising from  spontaneous breaking of scale invariance, i.e., dilaton, offers a simple and potentially elegant approach to the problem. In incorporating scale symmetry  to HLS to dense nuclear matter, in the past, the dilaton scalar was introduced as a lower-mass component of two-component structure of gluons figuring in the trace anomaly of QCD~\cite{br91,BR:DD}. The gluons that saturate the trace anomaly were decomposed into ``soft" and ``hard" components, with the soft-glue supposed to ``melt" at the chiral restoration, leaving the ``hard" component remaining above the chiral restoration point. The basic assumption was that the melting of the``soft-glue" can be tied to the bilinear quark condensate $\Sigma$ going to zero. Based on the behavior of the gluon condensate in temperature wherein roughly half of the gluon condensate vanishes at the critical temperature, it was assumed that the same ``melting" occurs in density at the critical density $n_c$. The CT approach does not rely on this separation. Instead it posits that there is an IR fixed point at large $\alpha$, say, at $\alpha_{IR}$ -- where $\alpha=g^2/4\pi$ with the $g$ the QCD gauge coupling -- at which the $\beta$ function goes to zero. At that point the trace anomaly $\theta^\mu_\mu=\frac{\beta(\alpha_s)}{4\alpha_s} G_{\mu\nu}^a G^{a\mu\nu} +(1+\gamma_m(\alpha_s))\sum_{q=u,d,s} m_q\bar{q}q\rightarrow \theta^\mu_\mu|_{IR}=\del^\mu D_\mu =(1+\gamma_m(\alpha_s))\sum_{q=u,d,s} m_q\bar{q}q$. At the IR fixed point and in the chiral limit with $m_q=0$, the dilatation current is conserved $\del^\mu D_\mu=0$ even if the gluon condensate and the quark condensate are non-zero separately or together. This combined symmetry is ``scale-chiral symmetry (SCS)." Spontaneous breaking of this SCS leads to a NG boson, dilaton, in addition to the pions (and kaons). The explicit breaking due to the trace anomaly and the quark mass makes dilaton a pseudo-NG.

As far as I know, there is no lattice evidence for such an IR fixed point for small number -- here 3 -- of flavors whereas there is evidence for it at large $N_f\sim 8$ as discussed in this conference by lattice experts. When asked whether such an IR fixed point for $N_f\sim 3$ is plausible, the general answer from lattice experts is ``not in matter-free space."  Although there is no dynamical lattice data, there is, however, a support for an IR fixed point for $N_f=2$ coming from numerical stochastic perturbation calculation at high fermion-loop order and Pad\'e approximation~\cite{stochastic}. While awaiting definitive confirmation/falsification of the existence of such an IR fixed point, I will simply assume it in the application to dense medium where such an IR fixed point could be envisioned to arise as an emergent phenomenon.

Now to implement this scale symmetry to (\ref{HLS}) to obtain scale-symmetric HLS (sHLS for short), I will use the standard trick with ``conformal compensator (or conformalon) field" with scale dimension 1,
\be
\chi=f e^{\sigma/f}
\ee
where $f$ is a spurion field for $f =\la\chi\ra$, the vev for a given vacuum (later defined in medium with density-modified ground state). The $\sigma$ represents the pNG scalar in nonlinear realization. What plays an important role in the CT theory is the anomalous dimension of the operator $G^2$ given by $\beta^\prime>0$, derivative of the $\beta$ function with respect to $\alpha_s$ near the IR fixed point. This corresponds to the slope with which the $\beta$ function approaches the IR fixed point and this can be large with positive sign. Due to this, writing down a fully general effective sHLS Lagrangian is rather involved and is not straightforward.\footnote{To illustrate what is involved, take the leading chiral order term in (\ref{HLS}), $\mathcal{L}_{(2)}^{\rm HLS}$. It has scale dimension 2, so the standard procedure of implementing scale symmetry would be to multiply it with $\bar{\chi}^2$ (where $\bar{\chi}\equiv \chi/f$). That would give a scale-invariant Lagrangian $\bar{\chi}^2\mathcal{L}_{(2)}^{\rm HLS}$. However due to the presence of $\beta^\prime$, one can have $c\bar{\chi}^2\mathcal{L}_{(2)}^{\rm HLS} +(1-c) \bar{\chi}^{2+\beta^\prime}\mathcal{L}_{(2)}^{\rm HLS}$ with an unknown constant $c$. Such a construction would make the resulting Lagrangian considerably more complicated with doubled number of parameters and hence unmanageable. To proceed however I will make the assumption that given that the dynamics of the matter fields that involve no scalar excitations is reliably given by (\ref{HLS}), $c$ could be taken to  be near 1. How the anomalous dimension of $G^2$ affects the matter field dynamics in various different observables remains to be investigated. It will be seen that in the presence of nuclear interactions in medium, there is locking of scale symmetry to chiral symmetry as soon as the matter density is non-zero. This suggests that $c\neq 0$ may have to be taken into account for a more realistic treatment. I will leave this matter for future work.}

Confining to the simplied possible form,  one can write
\be
\mathcal{L}_{\rm sHLS} & = & (\frac{\chi}{f_\sigma})^2
\mathcal{L}_{(2)}^{\rm HLS} + \mathcal{L}_{(4)}^{\rm HLS} +
\mathcal{L}_{4\rm anom}^{\rm HLS} +(\frac{\chi}{f_\sigma})^3\mathcal{L}_{\rm mass}+ V(\chi) +\cdots \, .
\label{sHLS}
\ee
Here $f_\sigma\equiv \la 0|\chi|0\ra$, vev in the matter-free vacuum (MFV for short) and $V(\chi)$ is the dilaton potential chosen so that together with the quark mass term, it is minimized at $\la\chi\ra=f_\sigma$. The ${\cal O}(p^4)$ terms (including the hWZ term) in (\ref{HLS}) are of scale dimension 4 and hence not multplied by the conformalons. Again the terms that introduce the anomalous dimension $\beta^\prime$ are ignored in both of them.

In what follows, I apply the simple Lagrangian (\ref{sHLS}) to baryonic matter and point out a few qualitatively striking features that could arise at high density.
\section{Scale-Chiral Symmetry in Medium}
\subsection{General structure}
I start with several general features that are known of the Lagrangian (\ref{sHLS}).
\vskip 0.2cm
$\bullet$ {Vector manifestation (VM) fixed point
\vskip 0.2cm
Let's first consider the chiral limit. Wilsonian-matched to QCD at a matching scale $\Lambda_M\lsim \Lambda_\chi\approx 4\pi f_\pi$, the Lagrangian (\ref{HLS}) has the VM fixed point at  which the $\rho$ mass goes to zero as~\cite{HY:PR}
\be
m_\rho|_{\Sigma\rightarrow 0} \sim g\sim \Sigma\rightarrow 0\, .\label{VM}
\ee
This is a feature that does not depend on how the VM fixed point is reached.
We would like to see whether high density does it and if so, how it does. Let me denote the critical density as $n_c$. The appropriate  value for $n_c$ -- thus far unknown in QCD -- required for making predictions will be specified later.

Next we need to have baryons in the system. We first treat baryons as solitons, i.e., skyrmions, arising from the Lagrangian (\ref{sHLS}). This is a natural way in large $N_c$ QCD. This endows the system with topology. That offers certain features of dense matter not manifestly visible in treatments with baryon fields explicitly incorporated into the theory. Dense bayonic system can be generated by putting skyrnions on crystal lattice and squeezing the crystal size to simulate density~\cite{multifacet}.  This approach has an advantage that both mesons and baryons, elementary as well as multi-body systems, are treated on the same footing. Unfortunately It  has the disadvantage that rigorous formulation is mathematically involved and for that reason has not been fully developed to enable one to confront experiments quantitatively. Some of the highly intricate issues are discussed in recent reviews~\cite{MPLB}. I am  not going to attempt this approach here. However I will exploit some of the robust features that topology provides in setting up the tools for doing EFT calculations.

The most important aspect of the skyrmion description for dense-matter physics is that at a density that we denote as $n_{1/2}$ that lies somewhat above the normal nuclear matter density $n_0\simeq 0.16$fm$^{-3}$,  the skyrmions in dense matter fractionize to half-skyrmions. This is a topological transition involving no local order parameters. When this happens, the chiral condensate $\Sigma$ which is related to the (bilinear) quark condensate in QCD language goes to zero {\it on average} within the unit cell of the crystal. The condensate however is not locally zero, so  in fact chiral symmetry remains spontaneously broken. The order parameter for this is therefore not the bilinear condensate $\la\bar{q}q\ra$ but presumably of multiquark condensates $\la(\bar{q}q)^n\ra$ for $n>1$ which are non-zero in the half-skyrmion phase. Thus the pion decay constant, which is usually associated with the quark condensate in continuum description, is not zero. The appearance of half-skyrmions with vanishing bilinear chiral condensate had been discussed before, but that chiral symmetry is not restored to  Wigner phase was first observed in \cite{pseudogap}. It turns out however that while chiral symmetry is not restored, parity-doubling takes place in the half-skyrmion phase~\cite{ma-pdoubling} \footnote{This is reminiscent of Georgi's vector symmetry where $\la\bar{q}q\ra=0$ with $f_\pi\neq 0$~\cite{georgi}.} with a {\it surprising consequence that some $\gsim 70\%$ of the proton mass (call it $M_0$), largely associated with gluon condensates, remain non-zero when the condensate $\Sigma$ averages to zero.} Thus a large portion of the proton mass is not due to the standard Nambu-Goldstone mechanism of chiral symmetry spontaneous breaking as was generally thought~\cite{schumacher}.  One can {\it approximately} identify $M_0$ as a chiral invariant mass in parity-doubled baryon models. What is equally significant is that the appearance of this half-skyrmion phase, in this formulation, is found to play the key role in the structure of the EoS for compact stars.
\vskip 0.2cm
$\bullet$ {Topological demarcation of density regimes R-I and R-II}
\vskip 0.2cm
The impact of the topological change at $n_{1/2}$ is found in the change in the structure of the Lagrangian for the density regions below and above $n_{1/2}$. It will be convenient to make the demarcation into the two density regions
\be
{\rm R(egion)-I}&:&\ \  0<n< n_{1/2}, \label{RI}\\
{\rm R(egion)-II}&:& \ \ n_{1/2}\leq n\leq n_c.\label{RII}
\ee
At present, the demarcation density $n_{1/2}$ is difficult to pin down precisely. However on phenomenological considerations and its robustness to dilaton properties such as mass and coupling constants, one can make an estimate. It ranges  $n_{1/2} \sim (1.5-2.0)n_0$.  It is found in numerical applications that  $n_{1/2}\sim 2.0n_0$ fits in well. Hopefully  this region will soon be accessed experimentally in future heavy-ion accelerators.

We will see that one can infer from scale-chiral symmetry considerations how the ``bare" parameters of the Lagrangian behave in the two regions.
\vskip 0.2cm
$\bullet$ Cusp in symmetry energy
\vskip 0.2cm
For EoS for compact-star matter, the important quantity is the energy per nucleon for matter with $P$ protons and $N$ neutrons, $E(n, x)$,
\be
E(n,x)=E(n,0)+Sx^2+\cdots\label{E}
\ee
with $x=(N-P)/(N+P)$.  Here $S$ is what is referred to as ``symmetry energy factor." The pure neutron matter is for $x=1$ and the symmetric nuclear matter is for $x=0$.

 The most striking feature characterizing the topological property of skyrmion matter observed on crystal is a cusp structure in $S$ at $n=n_{1/2}$~\cite{cusp}: $S$ drops as density approaches $n_{1/2}$ and then turns up and increases at $n_{1/2}$. One might wonder whether this feature is not just an artifact of the crystal structure. It has however been shown that it is not an artifact but a rather robust feature of the topology change and can be justified by a microscopic argument.

The  $S$ can be derived in the skyrmion description by collective-quantization of the skyrmion neutron matter. It is a semi-classical quantity. Yet it contains certain features of quantum many-body features of standard many-body approaches. For instance, this sort of ``classicalized" quantum effects are manifested in certain scattering processes computed in classical skyrmion processes~\cite{manton}. In fact, implementing the VM (\ref{VM}) in the Lagrangian (\ref{sHLS}), it has been shown that the change in the structure of the tensor forces at $n_{1/2}$ in quantum many-body calculation reproduces the far-from-obvious cusp structure~\cite{cusp}. Let me stress that this represents one of my main themes in the talk, that is, combining HLS with topology leads to a novel prediction not found in nontopological approaches. In terms of the nuclear tensor forces, the density $n_{1/2}$ sets the demarcation of different tensor force structures, below $n_{1/2}$ an intricate interplay of pion and $\rho$, the two compensating each other, and above $n_{1/2}$ the pion field taking over the tensor forces~\cite{LRtensorforces}.
\subsection{In-medium effective Lagrangian}
Let me focus on a Lagrangian of the form (\ref{sHLS}), the ``bare" parameters of which are Wilsonian-matched via correlators to QCD~\cite{HY:PR}. The matching is made at a matching scale commensurate with the chiral scale $4\pi f_\pi$\footnote{Whether this matching can be reliably done for the problem is not clear. Although calculations are feasible in principle, I won't specifically rely on the specific form obtained in \cite{HY:PR}.} but the Lagrangian can be brought down to a phenomenologically more relevant scale, say,  just above the $\rho$ mass scale in the matter-free space. The power of the matching is that it endows the parameters of the Lagrangian with nonperturbative quantities of QCD, typically of various condensates -- such as gluon, quark etc. condensates.  It turns out that the main condensates are the quark condensate $\la\bar{q}q\ra$, and the gluon condensate $\la G^2\ra$ but there are also various mixed condensates. The condensates, being the vacuum expectation values (vev) of  local operators,  reflect the vacuum structure. Therefore when the vacuum is modified by -- in our case temperature, density etc.--  the condensates will ``slide" with the complex vacuum. This implies the ``bare" parameters of the Lagrangian determined at the matching scale will depend on the vacuum structure, i.e., on density, in nuclear medium. We therefore implement this sliding vacuum structure in applying the Lagrangian to dense nuclear matter.

To take into account the vacuum property in the presence of baryonic matter, one then expands the scalar field
\be
\chi=\la 0^\star |\chi| 0^\star\ra+\chi^\prime
\ee
where $|0^\star\ra$ stands for the ground state with baryonic matter, say, $\star$-vacuum. Thus $f^\star_\sigma=\la 0^\star|\chi| 0^\star\ra$,  the in-medium decay constant ($f_\sigma=\la 0|\chi|0\ra$ is the decay constant in the matter-free vacuum). The expanded Lagrangian takes the form
\be
\mathcal{L}_{\rm sHLS} & = &({\frac{f^*_\sigma}{f_\sigma}})^2
\mathcal{L}_{(2)}^{\rm HLS} + \mathcal{L}_{(4)}^{\rm HLS} +
\mathcal{L}_{(4)\rm anom}^{\rm HLS} +({\frac{f^*_\sigma}{f_\sigma}})^3\mathcal{L}_{\rm mass}+\cdots \,.
\label{sHLS2}
\ee
The dilaton terms including $\chi^\prime$ coupling to pions, matter fields etc. are in the ellipsis and not shown explicitly.

In setting up the Lagrangian to be applied to dense matter, I will incorporate the nucleon fields explicitly into the Lagrangian (\ref{sHLS2}) in chiral-scale symmetric way~\cite{LPR15}. From now on, I will call it $bs$HLS Lagrangian.
\subsubsection{Scaling ``bare" parameters of effective Lagrangian}
Frst I discuss the structure of the Lagrangian applicable in R-I.

In terms of the topological structure given by the sHLS Lagrangian (\ref{sHLS2}), this region is populated by skyrmions. So it corresponds to normal skyrmion matter. This region is characterized by one density-scaling parameter
\be
\Phi (n)=f^\star_\sigma/f_\sigma.
\ee
The first observation that follows naturally from the locking of chiral symmetry and scale symmetry is that\footnote{I use the approximate relation instead of equality because as noted above, the Lagrangian (\ref{sHLS}) ignores possible effects of the anomalous dimension $\beta^\prime$.}
\be
f_\pi^*/f_\pi \approx\Phi
\ee
which leads to
\be
m_\pi^*/m_\pi\approx {\Phi}^{\frac 12}.
\ee
Up to nuclear matter density $n_0=0.16$fm$^{-3}$, $\Phi$ can be obtained from experiments (by looking at deeply bound pionic nuclear systems) or perhaps from chiral perturbation theory. I will assume that one can extend this up to $n_{1/2}\approx 2n_0$.

It also follows straightforwardly that
\be
m_N^*/m_N\approx m_\rho^*/m_\rho\approx m_\omega^*/m_\omega\approx m_\sigma^*/m_\sigma\approx \Phi\, ,
\ee
\be
g^*_{\sigma NN}/g_{\sigma NN}\approx g^*_{\omega NN}/g_{\omega NN}\approx g^*_{\rho NN}/g_{\rho NN}\approx 1
\ee
and
\be
g_A^*/g_A\approx g^*_{\pi NN}/g_{\pi NN}\approx \Phi\, .
\ee
Far away from the critical density $n_c$, one can parameterize
\be
\Phi(n)\approx \frac{1}{1+c_I n/n_0}\label{Phi}
\ee
with $c_I<1$.

In sum, one constant of the scaling parameter $c_I$ -- which can be connected to data on deeply bound pionic nuclear systems -- completely specifies the ``bare" Lagrangian in R-I with which one can do EFT calculations in medium up to density $n_{1/2}$.
So the physics around -- and in the vicinity above -- $n_0$ is given almost entirely by the one parameter $c_I$ which is controlled at least up to $n_0$ by available data.

The situation in R-II is much less clear. There is no direct help from QCD, i.e., no lattice QCD nor any reliable theoretical tools at high density. In skyrmon picture, due to the topology change at $n_{1/2}$ from skyrmions to half-skyrmions in R-II, when translated to the structure of the effective Lagrangian, $bs$HLS, the ``bare" parameters will undergo drastic modification. I will first infer the structure of the parameters from what we know from the skyrmion-crystal simulation.

The first thing to note is that the nucleon mass in R-II goes to a constant $M_0$ when one writes the bare parameter for the nucleon mass as
\be
m^*_N=M_0 +\Delta (\Sigma)\label{Nmass}
\ee
with  $\Sigma\propto \la\bar{q}q\ra$ going to zero approaching $n_c$  in R-II. As noted, although $\Sigma=0$ in the half-skyrmion phase, the pion is still excited with a nonzero pion decay constant. This means that chiral symmetry is in the NG mode, with massive hadrons, that are parity-doubled. This is seen in the skyrmion crystal model with (\ref{sHLS})~\cite{maetal,ma-pdoubling}. This is reproduced also in a renormalization-group analysist of hidden local symmetric parity-doublet model with dilaton limit fixed point (DLFP)~\cite{paengetal}. In both cases, the ``nearly" chiral invariant mass\footnote{Note that in the skyrmion crystal formulation, $\Sigma$ averages to zero in a unit cell but chiral symmetry is still broken. However chiral symmetry breaking is characterized by highly suppressed multi-quark condensates, with strongly reduced pion mass.} $M_0$ comes out to be $M_0\approx (0.7-0.8)m_N$. I will therefore take $m^*_N\approx \kappa m_N$ with $\kappa\approx M_0/m_N$ in R-II.

I should point out that QCD lattice calculations at high density for non-zero $\beta$ in strong coupling expansion {\it do} provide a support for parity-doubling in the confined phase~\cite{tomboulis}.

Both the skyrmion crystal and mean-field approaches~\cite{maetal,paengetal} lead to the relation $f^*_\sigma/f_\sigma\approx f^*_\pi/f_\pi\approx m^*_N/m_N$. As stressed in \cite{LPR15}, this relation together with $m^*_\sigma/m_\sigma\approx f^*_\sigma/f_\sigma$ follows from low-energy theorems for chiral-scale symmetry as applied in medium. Thus I take
\be
f^*_\sigma/f_\sigma\approx f^*_\pi/f_\pi\approx m^*_N/m_N\approx m^*_\sigma/m_\sigma\approx \kappa.
\ee

I now turn to the vector mesons.

For the $\rho$ meson, I will assume for numerics that the VM fixed point comes at $n_c\approx (6-7)n_0$. Considering that R-II is a density regime ``closer" to the VM fixed point, I will simply take  from Eq.~(\ref{VM}) what could be applicable for the whole range $n_{1/2}\lsim  n <n_c$ of densities,
\be
m^*_\rho/m_\rho\approx g^*_{\rho NN}/g_{\rho NN}\approx (1-n/n_c).
\ee
This is undoubtedly a drastic simplification. For instance, near $n_{1/2}$ which is not very close to $n_c$, the density dependence could be more involved. Nonetheless I think it captures the essential feature of the process.

Now the $\omega$ meson. As noted, the $U(2)$ symmetry may be applicable in R-I, so
\be
m^*_\omega/m_\omega\approx \Phi,
\ee
the same as (\ref{Phi}).  However the properties of the $\omega$ mass and $\omega$-nucleon coupling are unknown in the whole region of R-II. These constitute the main uncertainty in the effective $bs$HLS Lagrangian. One can however make some progress from the present understanding of what's going on in massive compact stars.

What is most significant in the present framework is that once the symmetry energy factor $S$ is suitably  controlled by the pion and $\rho$ properties, i.e.,  through the net tensor force that dominates the factor $S$ with very little effect from $\omega$ and $\sigma$, it is the interplay between the dilaton and the $\omega$ meson in the symmetric ($x=0$) part of the energy (\ref{E}) that plays the most important role in the EoS of dense compact star medium. One can easily see what happens by looking at the potential energies associated with the dilaton and $\omega$ exchange which have the form \{$-(\frac{g^*_{\sigma NN}}{m^*_\sigma})^2 n_s+ (\frac{g^*_{\omega NN}}{m^*_\omega})^2 n_B$\} where $n_{s}(n_B)$ is the scalar (baryon number) density.  Then it is the competition between the dilaton attraction and the $\omega$ repulsion that controls the game. The ratio $Y_\sigma\equiv (\frac{g^*_{\sigma NN}}{m^*_\sigma})^2$ is roughly constant without scaling whereas the ratio $Y_\omega\equiv (\frac{g^*_{\omega NN}}{m^*_\omega})^2$ is likely to scale in density. Taking into account that $n_B\gsim n_s$ for $m^*_N < \infty$, given that in equilibrium nuclear matter, the two roughly cancel giving rise to a small binding energy observed in nature, namely,  a BPS structure in the language of skyrmion, one can see that the repulsion could overcome the attraction at some density above $n_0$ with a dropping $\omega$ mass. This repulsion is needed to resolve the strangeness (hyperons and kaon condensation) problem and to account for the massive $\sim 2$ solar mass neutron stars.

How the repulsion overcomes the attraction is constrained by two crucial factors: one is that the sound velocity of the matter not violate causality in the density regime relevant for the stars and the other is that the EoS not become too softened by the appearance of the hyperons or kaon condensations. The latter will obstruct the formation of the observed massive ($\gsim 2$-solar mass) stars. Both can be avoided by tuning the repulsion due to the scaling $\omega$ mass. What is found necessary is that the $\omega$-NN coupling $g_{\omega NN}$ and the $\omega$ mass $m_\omega$ scale slowly in density in the same way $\sim m^*_\omega /m_\omega\approx h g^*_{\omega NN}/g_{\omega NN}$ with $h<1$.

The effective $bs$HLS Lagrangian so constructed sliding in density has been applied to compact stars with rather satisfactory results~\cite{PKLR}. Given the space limit for this note, I will skip the details here and mention merely that with only 3 ``bare" parameters, 1 in R-I ($c_I$) and 2 in R-II ($\kappa$ and $h$) with some fine-tuning -- presumably associated with $1/N_c$ corrections -- within the range of parameters obtained, both normal nuclear matter and massive compact stars come out in consistency with Nature.

\section{Conclusions}
In the description that combines hidden local symmetry of the $\rho$ vector meson and scale invariance of the dilaton scalar $\sigma$, a large portion of the proton mass $M_0\sim 0.7m_N$  is found to remain non-vanishing in the limit that the quark condensate, dialled by density, goes to zero in the chiral limit. The non-vanishing mass $M_0$ is chiral invariant, hence implies parity doubling. The source for $M_0$ then cannot be accounted for by the standard Nambu-Goldstone mechanism anchored on spontaneous breaking of chiral symmetry for mass generation. It is in a sense  ``mass without mass"~\cite{wilczek}, perhaps lying  outside of the Higgs paradigm. This picture is arrived at when scale invariance associated with an as yet unproven IR fixed point in QCD gauge coupling is implemented to hidden local symmetry unified into a combined chiral-scale symmetry. A striking consequence of this picture is the structure of the EoS of compact star matter accommodating the massive neutron stars recently observed that follows from topological transition of matter from skyrmions to half-skyrmions predicted at large $N_c$ of QCD. While QCD proper is incapable of addressing the high density regime involved, Nature seems to indicate rather unambiguously that the $\omega$ meson is {\it not} on the trajectory that leads to the VM fixed point of the $\rho$ meson, signalling how $U(2)$ HLS for the light vector mesons could be breaking down. As far as I know, this is the first hint from Nature that the $\omega$ meson may not behave the same as the $\rho$ nearing chiral restoration.

How the intricate locking of chiral symmetry and scale symmetry and the unbreaking of mended symmetries~\cite{weinberg} involving pseudo-Nambu-Goldstone bosons -- pions, kaons and dilaton ($\sigma$) -- and vector mesons -- $\rho$, $\omega$ and $a_1$ -- remain an open issue in nuclear physics as well as in astrophysics.
\subsection*{Acknowledgments}
The work described here is based on works done largely in collaboration with Masa Harada, Tom Kuo, Hyun Kyu Lee, Yong-Liang Ma and Won-Gi Paeng. I would like to thank them for  discussions and help. I am particularly grateful for encouragements and support of Koichi Yamawaki and of course for his invitation to this meeting.

\end{document}